\documentclass[conference]{IEEEtran}
\IEEEoverridecommandlockouts
\usepackage{cite}
\usepackage{amsmath,amssymb,amsfonts}
\usepackage{algorithmic}
\usepackage{subcaption}
\usepackage{graphicx}
\usepackage{array}
\usepackage{caption}
\usepackage{graphicx}
\usepackage{verbatim}
\usepackage{url} 
\usepackage{textcomp}
\usepackage{xcolor}
\usepackage{stfloats}
\def\BibTeX{{\rm B\kern-.05em{\sc i\kern-.025em b}\kern-.08em

    T\kern-.1667em\lower.7ex\hbox{E}\kern-.125emX}}
\begin{document}

\title{Power Grid Infrastructure for
AI Data Centers}

%\author{Authors}
\author{
Amir Sajadi,  \IEEEmembership{Senior Member, IEEE}, 
Muhy~E.~Za'ter, \IEEEmembership{Student Member, IEEE}, 
Maria Vabson, \\
Kyri Baker, \IEEEmembership{Senior Member, IEEE}, and 
Bri-Mathias Hodge,  \IEEEmembership{Senior Member, IEEE}

\thanks{Authors are with the University of Colorado Boulder, Boulder, CO 80309, USA, Email: \{Amir.Sajadi, Muhy.Zater, Maria.Vabson, Kyri, BriMathias.Hodge\}@colorado.edu.} 

}

%\markboth{ {\color{red} WORKING DRAFT}, 25th, September~2025}%
%{Sajadi \MakeLowercase{\textit{et al.}}: Power Grid Considerations for AI Data Centers}

\maketitle
\thispagestyle{plain}
\pagestyle{plain}

\begin{abstract}

This article addresses recent advances in artificial intelligence, which have set off an astounding race among technology frontiers to build large data centers. It provides insights into impacts of large data centers on the planning and operation of the power grid.

\end{abstract}


\section{Introduction}

The field of artificial intelligence (AI), particularly large language models (LLMs), has taken the world
by storm in recent years. Fueled by recent advances in deep learning and massive datasets, these models have demonstrated remarkable abilities across a broad range of applications and a strong potential for positive contributions in many economic sectors. Consequently, a fierce competition has been sparked among technology leaders, and even nations, in the development of large-scale data centers, creating a global surge in electricity demand. 

Data centers and chip foundries are projected to grow exponentially in the coming years. Meeting their rising electricity demand will require substantial investment to expand and reinforce power grid infrastructure. However, upgrading this infrastructure is a lengthy process that can take several years and often leads to long interconnection and commissioning timelines for data centers, particularly for large data centers, mainly due to constrained power availability \cite{data_center_how_long_to_build}. As an emergency remedy, many utilities are taking near-term action to alleviate the strain on their assets. For example, some utilities reversed previous positions on the retirement of some coal units to operate beyond their originally planned lifetimes and proposed the construction of new fossil fuel units. Similarly, some nuclear plants are being brought out of retirement, and more broadly, data centers are single-handedly driving a lot of new nuclear development that was not seen before. The mounting pressure from the widespread construction of large data centers leaves many power utilities at the perils of aging infrastructure and rising complexity of grid operation, in addition to challenges around system planning and upgrade needs to meet the reliability and affordability standards. 

Furthermore, the inherent electric characteristics of these data centers and their unconventional load shape and size for both training and inference in LLMs introduce complexity to the power grid planning and operation. For instance, during training, tens of thousands of processors can instantaneously spike power consumption up or down due to factors such as waiting for checkpointing or collective communication to complete or the startup and shutdown of entire training jobs. This stands in contrast with the conventional moderate ramp load profiles that power grids are designed to routinely manage. Such events can introduce power quality challenges \cite{sun2019modeling} and can cause sudden fluctuations in data center power usage, sometimes reaching hundreds of kilowatts (kW) to megawatts (MW) \cite{dubey2024llama}. These extreme power ramp rates are unprecedented in large industrial loads, especially since they occur within millisecond timescales. In contrast, inference is heavily influenced by user demand for a given LLM, leading to potential surges at peak times. 

Many of the grid critical challenges and complexities with regards to large data centers have been documented in the literature with technical analysis, use cases, and case studies \cite{chen_new}~--\cite{PES_report}. This article examines the risks and opportunities that large AI data centers present, given contemporary electric power industry processes and practices. The topics discussed are organized into two main categories:

\begin{enumerate}
	\item Grid Expansion and Market Strategy
	\item Grid Performance and Reliability
\end{enumerate} 
First, we introduce the key components inside AI data centers and their power consumption profiles. Subsequently, we discuss the integration of data centers into the power grid and explore the associated challenges. For each topic, we highlight the emerging trends and issues on the horizon, along with recommendations as appropriate, intended for policy makers, electric utilities, and data center developers and operators.

\section{Overview of Large Data Centers}

\subsection{Data Center Components}

A typical data center consists of many servers, ranging from 500 to 2,000 for small data centers to 2,000 to 10,000 servers for medium data centers and numbering above 10,000 servers for hyperscale data centers \cite{nescoeDataCenters}. Each server consists of computation units, which are the primary consumers of power in a data center. These can be either AI-accelerated graphics processing units (GPUs) and tensor processing units (TPUs) or non AI-accelerated central processing unit (CPUs). The CPU has been the standard technology of computation and is widely used in personal computers and professional servers. On the other hand, GPUs and TPUs are emerging processing technologies that consume more power than CPUs, but their higher processing speeds are valuable for AI applications in large data centers. Additionally, each server consists of storage and network units for data storage and communication whose energy consumption relative to that of the processors is much lower. Processing units, when stacked up in a large data center and operating at full load, can sum to large quantities of electric power, on the order of MW and expected to rise to gigawatts (GW). The energy demand from the IT components in AI-accelerated servers alone, accounting for all processing, storage, and communication and network equipment, is estimated to contribute between 50\% and 70\% of total data center energy consumption \cite{shehabi20242024}.

The second most energy consuming element of a data center is its cooling system. Its electricity consumption is determined based on the scale of the data center, the type of cooling system, and the efficiency of the technology used, as well as the data center location due to additional ambient cooling needs, especially in hotter or more humid climates. Cooling typically accounts for up to 40\% of the total electricity consumption of a data center \cite{aljbour2024powering}. 

The third energy consuming element is the lighting system. This energy consumption can be readily calculated based on power usage of the lighting technology used (commonly LED) and the number of lights used in a facility. The final major energy consuming component is the electric losses in the electricity distribution network of the facility, which can account for 10\%–12\% of the total power  \cite{energystarReduceEnergy}. This can be reduced by utilizing direct current (DC) technology and operating at higher voltage levels. 

Power usage effectiveness (PUE) is the most common metric used to quantify the amount of electric energy a data center consumes for both IT and non-IT components. It is the ratio of the total power consumed by a data center (including IT load, cooling system, lights, and electricity losses) to the power consumed by the computer systems (which is the total aggregate of all power readings from each server rack). In most modern efficient data centers, especially hyperscale and AI data centers, the PUE is around 1.05–1.15 (the
lower and closer to 1.0, the better) \cite{verdecchia2023systematic}, whereas in conventional data centers, this number can rise to 1.6.

\subsection{Data Centers Scale in Numbers}

In 2023, data centers in the United States consumed approximately 4.4\% of the country’s total energy consumption. This demand is projected to rise significantly in the near future, reaching 6.7\% to 12\% of nationwide electricity consumption by 2028  \cite{aljbour2024powering}. A major driver of this growing energy demand is the increasing deployment of AI-accelerated computing, particularly GPU-based servers, which are significantly increasing in both power consumption and volume. In 2023, the average power rating per AI server instance (typically consisting of eight GPUs) was approximately 8.5 kW, with operational power averaging around 6 kW. By 2028, these values are expected to increase, with power ratings reaching 11 kW and operational power ranging between 6.5 and 8.5 kW \cite{shehabi20242024}. Even when idle, servers consume about 20\%–25\% of their peak power, contributing to overall energy inefficiency \cite{wang2023overview}.

As AI usage continues to scale up, the number of installed servers in the United States is projected to reach 37 million by 2028, with AI-specific servers accounting for 8 to 12 million of them \cite{shehabi20242024}. This growth is driven by the surge in use of AI products, which increases utilization rates. Training instances currently operate at around 80\% utilization and are projected to rise to 90\% by 2028 \cite{epoch2024trainingcomputeoffrontieraimodelsgrowsby45xperyear}, while inference instances, which currently utilize about 40\% of their capacity, are expected to reach 60\% \cite{epoch2024optimallyallocatingcomputebetweeninferenceandtraining}. In addition to the energy-intensive nature of the processing units, other components, such as storage and networking also contribute to power consumption. Also, AI-focused data centers utilize efficient cooling technologies that typically have a lower PUE, indicating relatively efficient power management, although sustainability concerns persist.

\subsection{AI Load Profile and Characteristics}

A key contributor to this surge in electricity demand is the training of large-scale AI models, and more specifically LLMs, which require vast computational resources. For instance, the training of Meta’s LLaMA-3 405B model was conducted on 16,000 H100 GPUs, distributed across instances of eight GPUs per server \cite{dubey2024llama}. Similarly, Google’s PaLM model and DeepSeek’s latest model are on the order of thousands of processing units.

Power consumption associated with AI workloads can be broadly categorized into two primary types: 1) training and 2) inference. Training includes both the pretraining and fine-tuning stages of model development, each of which demands substantial computational resources. The power requirements for training are exceptionally high, often involving thousands of high-performance GPUs operating near full capacity over extended periods, sometimes spanning days or even months. Although training is not entirely dispatchable, it demonstrates a degree of flexibility in scheduling. Specifically, training processes can be paused and resumed later without significant issues as they are neither real time nor highly latency sensitive. However, instantaneous stopping is not feasible without the risk of losing unsaved progress. A characteristic of LLM training is the abrupt power load variation that occurs during checkpoint saving, where the load can rapidly drop from full utilization to near idle within seconds.

On the other hand, inference, while requiring considerably lower power than training due to reduced computational intensity, presents different operational challenges. Unlike training, inference workloads are primarily driven by user demand, making them less predictable and less flexible. Furthermore, inference is a real-time, latency-sensitive process, meaning that it must respond instantaneously to input requests without significant delays. Consequently, it is a nondispatchable load if served synchronously as its execution cannot be deferred or paused without degrading the quality of service. The unpredictability of inference demand, coupled with its real-time nature, may pose significant challenges for the management of the host power grid. As a potential remedial solution, some commercial LLM providers such as OpenAI and Google have recently begun developing mitigation solutions to reduce stress on the grid by offering asynchronous inference for load flexibility. However, the underlying challenge remains unresolved and continues to represent a critical source of uncertainty to the grid planners and operators.

\begin{figure*}[!]
	\centering
	\includegraphics[width=2\columnwidth]{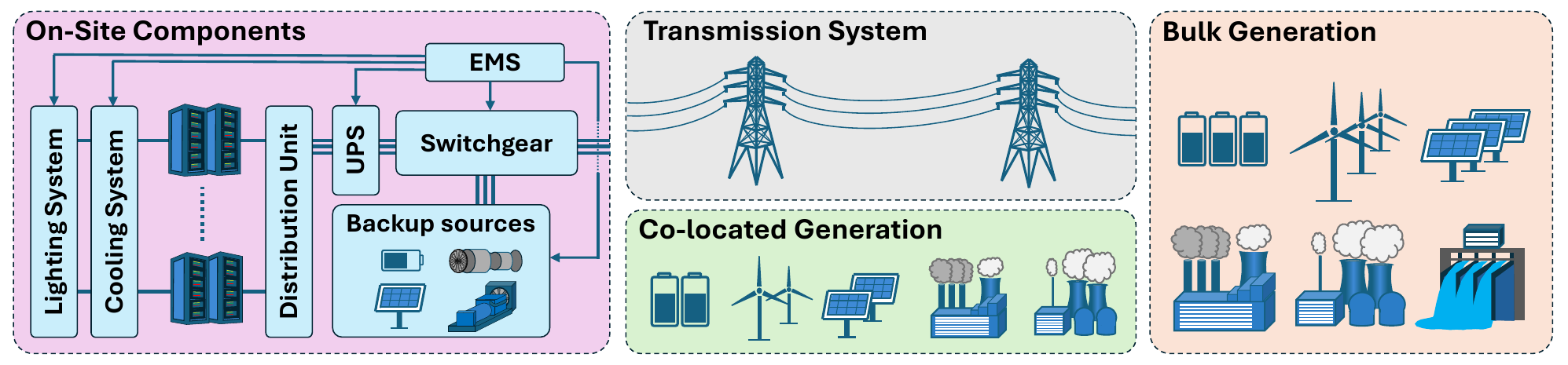}
	\caption{Data centers as part of the power grid ecosystem. EMS: energy management system; UPS: uninterruptible power supply.}
	\label{fig:grid_schematics}
\end{figure*}

\subsection{Interconnection to Power Grid}

The interconnection of AI data centers to the electric power grid creates tight interactions and potentially symbiotic relationships between their respective systems and subsystems, as conceptualized in Figure \ref{fig:grid_schematics}. As a result, close coordination is essential to prevent adverse impacts on grid components and to ensure safe integration of these data centers. The remainder of this article examines key issues related to their interconnection, associated implications, and important planning and operational considerations.

\section{Grid Expansion and Market Strategy}

\subsection{Load Growth Forecast}

A core challenge with grid planning for  data  center  interconnection  is accurately forecasting load growth driven by expected data center additions. The interconnection of large loads on its own has a long precedent in the power system industry. However, two barriers distinguish the data centers from the existing convention of forecasting large loads: 1) the lack of easily accessible data regarding power consumption in data centers and their specific patterns and 2) the lack of easily accessible development plans that are often treated as trade secrets. A solution to remove the former barrier would be to make electric power consumption data of these facilities available to the grid planners. This type of data with hourly resolution (to represent daily, seasonal, and annual variations and to capture variations throughout the day, including ramps, peaks, and unique shapes) for multiple years would be sufficient to inform the power system planners. Additionally, data on daily peak and average power consumption would be helpful to determine power generation and transfer capacity needs and the cost of energy, respectively. Such data could be either provided from historical measurements or synthesized. The use of historical measurement would be beneficial to extrapolate future scenarios with a more limited uncertainty and obtain accurate hourly profiles. On the other hand, the development and use of synthetic data would require detailed information about the design and topology of the data center and the technology planned to be employed. As such, it might need a wider range of future scenarios to be investigated to account for a greater uncertainty which is embedded in the synthetic data. 

The latter barrier could be lowered by putting in place a framework where the data center developers have a statutory timeline to publicly disclose their construction or expansion intent and plans. This timeline and the notice periods can be determined by the regulators and subject to the needs of the hosting electric utility. It would give the planners sufficient time to prepare the grid to reliably serve the data center without compromising the reliability of energy delivery to its ratepayers. Other efforts such as putting in place financial incentives like better electricity rates or tax breaks could encourage the data center developers to disclose their plans reasonably in advance

\subsection{Resource Adequacy and Capacity Expansion}

The rapid growth of large data centers challenges the availability of sufficient generation  capacity  with  adequate ramping rates on three fronts. They are 1) the time for permitting and constructing new generation facilities, 2) the possibility of introducing acute power demand fluctuations, and 3) the interconnection timelines. First, constructing new generation units can be a lengthy process, depending on the type of generation facility. In the United States, the average time for building most types of new generation facilities is more than two years, whereas a data center can be built and commissioned in less than two years. For example, while solar has the shortest time for implementation, one to two years, it is highly location dependent due to the amount of solar energy varying locations receive and the space requirements needed for sufficient capacity, which is a concern for scalability of large data center sites. Similarly, there are additional considerations for the other power plants in terms of where they can be both better placed for new construction and relicensed for prolonged use.

Second, large data centers can introduce significantly acute power demand fluctuations throughout the day in the form of high-power pulsated ramp up and down as their power consumption varies by their computation queries. Two approaches exist to address this issue. The first approach is to equip the grid with more fast-ramping backup units, which presents the risk of suboptimal planning or overbuilt. Such support was traditionally provided using fast ramping gas turbines and coal units, which are not the most environmentally friendly options. Also, conventionally, hydropower plants have been used for fast-regulation purposes, which could still be viewed as a more environmentally friendly option. While construction of new hydropower plants remains restricted to specific geographical locations and water availability, there could be more focus on relicensing existing facilities. It should be noted that the ramp rate of all conventional resources, even those that are traditionally recognized as fast-ramping units, is much slower than that of data centers. More recently, inverter-based resources, which consist of utility-scale batteries and variable renewable units (when available and operating with a headroom), offer a greener and ubiquitous solution with capabilities to more closely match the ramp rates of data centers. Regardless of the generation technology, the addition of more generation units always hinges on the availability of transmission capability, which adds a layer of interdependency to the feasibility of this approach. The second approach is to equip the data center facility with resources to smoothen its fluctuations and limit its ramps. The latter approach could transform the data center into a self-regulating load and could be achieved via options such as on-site diesel, gas generators, batteries, or other energy storage systems, subject to the local/regional environmental restrictions and air quality requirements and depending on the developer’s appetite for reducing the carbon footprint of their facilities considering the higher cost of most storage units. The choice of using battery storage as a backup resource, however, adds an extra layer of complexity to system planning as a new question arises about the optimal sizing for battery storage units, including both short-term battery and long-duration storage systems. This question could be multifaceted to understand the adequate size of the battery to 1) minimize (ideally eliminate) the fluctuations that the data center introduces to grid and 2) maximize the value of power generation from solar and wind units. We note that the small modular reactors and fuel cell industries have gained substantial interest in this space. Despite not yet having reached commercial maturity, both technologies could play a major role in the future. To address this problem, in the long term, we suggest that energy regulators may consider mandating data centers to self-regulate their fluctuations and be allowed on the network only if they satisfy certain ramp limits.

Third, the interconnection of new generation is a lengthy process and can take years. New generation to supply a modern AI data center will likely be a project greater than 20 MW and will take more than two years\cite{rand2024queued}. Between 2015 and 2020, the average timeline for the construction and commissioning of large data centers was reported to be around three years, with as little construction time as a year \cite{data_center_how_long_to_build}. However, in recent years, this timeline has increased to up to six years, mainly due to the lack of power infrastructure \cite{data_center_how_long_to_build}. Moving forward, the timeline gap between power plants and data center constructions needs to be eliminated to achieve sustainable growth. 

A common practice for data centers to avoid delays caused by the generation interconnection process is to develop or utilize a generation facility near the data center and directly connect the two facilities, therefore bypassing the grid. Nevertheless, a colocated generation facility could be still physically connected to the main utility grid with limited power exchanges and certain obligations and regulations, which are not clearly defined. This is an area where regulators need to pay closer attention and ease the path for developers.

\subsection{Transmission System Upgrades}

Large data centers are emerging all across the United States, and it is safe to suggest that they are pushing transmission systems to their limits. Therefore, increasing the transmission capacity across the country is the key to ensuring the reliability of the grid with adequate redundancy in the years to come.

Similar to the process for generation capacity expansion, the transmission system and associated substation construction and upgrades can be a lengthy process and takes several years, whereas a large data center, as noted earlier, can be constructed and commissioned in less than two years. This timescale discrepancy between the demand growth and network upgrade can delay the interconnection of data centers and, thus, result in a loss of revenue for the utilities.

Of the available options to increase the ampacity of a transmission line, reconductoring using advanced conductors presents a viable option to increasing its carry capacity (double to triple) with minimal modification to the transmission towers. Nonetheless, their deployment is costly and needs to be carefully studied to ensure it is economically justified. 

At the heart of transmission system planning is cost allocation, which determines the mechanism for investment recovery and return on equity. One may rely on intuition to suggest that if the line upgrade is required because of interconnection of a large data center as the main beneficiary, then the “beneficiary pays.” But defining beneficiaries for cost allocation purposes is often murky and not straightforward. Notwithstanding that each regional transmission organization has an established set of regional and interregional transmission planning rules, cost allocation often serves as a source of dispute among stakeholders and remains a complicated topic, especially if it involves other loads or multiple data centers.

\subsection{Integrated Resource Planning}

The rapid pace at which data centers are coming online has compelled many utilities to take swift action to address their near-term risks. However, these facilities have long-term broader impacts well beyond the power grid as they could alter their local and larger communities, ranging from climate, environment, and individual health to socioeconomic consequences. Therefore, the development of these facilities needs holistic examination of all relevant parameters. This could be achieved through integrated resource planning (IRP), which aims to optimize the whole system in the long term and shed light on all direct and indirect costs of the data centers.

An effective IRP could be achieved through direct engagement of the local community in all stages of planning, construction, and operation to directly hear from them. Their engagement also helps address their concerns early on and to ensure the developments and their operation align with the community, local, and regional interests and standards. Their concerns may include carbon and water footprint, air and water pollution and mitigation efforts, electricity and water availability and prices, job creation and the local economy, and the impact on the housing market. Ongoing dialogue and iterative processes that include meaningful debates and open communication with the residents and community leaders are the key to continuously keeping the locals informed and engaged.

\subsection{Risk Management and Market Strategy}

All data center development projects are subject to risks of delays or failure due to a range of issues such as supply chain challenges, technical flaws, financial security, permitting, land acquisition, changes in the political or social landscape, and legal challenges. In the power industry, the rules of wholesale energy markets are well established for delays or withdrawal of generation or transmission projects. However, these rules need to be expanded to include risk management for large data centers. Rules should also be established to discourage and prevent data center withdrawal and define mechanisms for recovery of all network upgrade studies and asset costs incurred. Such expansions will hedge the reliability risks for system planners, the investment risks for developers, and the risks for public ratepayers.

Large data centers are relatively new in the power industry, and therefore, there is a lack of adequate data and sufficient knowledge about their behavior and performance. From the operations perspective, accessing data centers’ behavior and profile data can help the host utility with its multiday ahead and the day ahead load forecasts and improve its market strategy. This is a critical function of operational planning and determines the energy (electricity and/or fuel) procurement and bidding strategy in the market, directly impacting revenues/losses and the settlements. From the planning perspective, the creation of a reporting system for large data centers can help the grid planners better learn about these new assets and improve their processes and procedures. This reporting can contain information about the availability and performance of these facilities, efficiency performance, their involuntary disconnection from the grid, duration, cause, and values like PUE and water usage effectiveness. These reports can be collected and processed by the existing bodies, for example, the North American Electric Reliability Corporation or the Energy Information Administration, as a new reporting requirement, for example, the large-load availability data system.

Lastly, the on-site resources of large data centers, generators or battery storage units, can offer them some operational flexibility, particularly during the grid constraint periods. Such capability may help them expedite their interconnection to the grid, subject to the flexibility parameters. Additionally, we envision that in the future, the data center operator could use these resources to offer a range of services to support their host grid, either as a participant in the wholesale energy market, a direct service agreement, or a combination of both. These services will be discussed later in this article. From the market strategy perspective, the key enabler for their engagement in wholesale energy markets is having market rules and tariffs that provide financial incentives to data centers to participate. Achieving this regulatory milestone requires direct engagement of electric power utilities, energy commissions, and data center developers and operators.

\section{Grid Performance and Reliability}

\subsection{System Stability and Power Quality}

Interconnecting large data centers to a power grid will require network impact studies to demonstrate that the incoming data center does not have adverse impacts on the extant network. These loads are driven by an opaque concentration of electronic and digital control systems and can be pulsated, withdrawing large quantities of power from the network instantaneously without any or with a negligible amount of directly coupled mechanical inertia. Furthermore, any electronic equipment could easily introduce a large degree of nonlinearity, in addition to power pulses, which are induced by data/computation queries, collectively contributing to stability and power quality concerns. 

The adverse impacts of data centers on the grid performance can manifest themselves in steady-state, oscillations, or nonideal sinusoidal waveforms, implicating system frequency, nodal voltages, and phase angles. Steady-state stability concerns slower dynamics, in which gradually accumulating deviations from scheduled values or standard limits can eventually render the system infeasible or inoperable. The oscillations are faster dynamics and can be either sustained (forced oscillations) or temporary (for example, local or inter-area oscillations). These oscillations, if left unmitigated, can damage equipment and incur maloperations and may lead to forced outages with the potential to cause blackouts. The nonideal waveforms, on the other hand, contain higherfrequency components, known as harmonics, which could be integer multiples, subharmonics, or interharmonics of the nominal frequency and could also damage equipment, lower their life span, and force malfunctions. These behaviors should be closely studied and routinely monitored to ensure the data center’s interactions with other assets are safe and in compliance with the grid codes and regional regulation on harmonic limits, power fluctuations, and response obligations.

From the grid dynamics perspective, the large data centers could be the source of disturbances to the grid. On that note, these data centers can introduce a new class of contingencies that involve a rapid, significant drop in the amount of power consumption by load at a specific node or complete disconnection from the network (load tripping). The system planners should consider studying this class of contingencies from both the small- and large-signal stability perspectives, and operators should put in place appropriate remedial actions.

When an area is identified to suffer from an adverse impact caused by a planned large data center, then it is inevitable that network upgrades or reinforcement are needed as a mitigation plan. Traditionally, electromechanical solutions, such as synchronous condensers, and power electronics-based solutions, such as the static synchronous compensator and static VAR compensator, have been successful in improving a system’s dynamics and power quality. More recently, grid-forming inverters have reached the maturity needed for utility-scale deployment, and solid-state transformers are expected to reach a similar level within the next few years. Both technologies offer promising solutions for data center applications. 

An outstanding challenge in this domain is the lack of reliable, high-fidelity, industry-grade models to accurately represent their transients and power quality issues and efficiently allow for system-wide phasor and electromagnetic transient simulations. Access to such models and simulation capabilities are essential for the utility planners to adequately understand and describe all hardware and software components as they relate to grid interactions. To address this challenge, data center developers and power system planners need to work together to create generic, open access, industry standard dynamic models and practices for model verification and quality testing.

\subsection{Fault Response and Fault Ride-Through}

Short circuit fault response and ride-through capabilities are preventative measures against undesired frequency and voltage phenomena. The power industry is realizing that these grid-supporting functions may not remain solely on the generation side and could be enforced on the load side, especially for the large data centers. This need came to light following a data center disconnection in 2024, where a fault on a 345-kV line resulted in near-instantaneous disconnection of 1.5 GW of data center load due to voltage sensitivity \cite{NERC_load_incident_2025}. Load drop of such magnitude following a high-voltage fault was unprecedented and concerning.

The preventative measures could be the standardization of fault response from large data centers in the form of short circuit contribution and mandatory enforcement of ride-through capability. For these measures, data centers may rely on their onsite resources or other existing solutions. Most notably, synchronous condensers are widely used for increasing fault current and system strength as they produce short circuit current and prevent significant voltage depression.

The specific level of fault current contribution and requirements for ride-through can be determined by either the host utility or the regional reliability entity and included in the interconnection agreement as operational requirements. Mandating a contribution to short circuit fault current in the context of data centers not only helps with easier detection of fault conditions, but it serves to require data centers to stay online and not disconnect immediately.

\subsection{Data Centers as Grid Flexibility Assets }

AI data centers can range from tens of MW- to GW-scale loads; however, their software defined workloads and the redundancy in their on-site equipment make them uniquely capable of acting as potentially controllable grid assets. There are multiple sources of dispatchable flexibility that could be used to support the grid, including but not limited to the following:

\begin{itemize}
	\item Uninterruptible power supply (UPS) batteries: UPSs regulate power and provide clean AC waveform to the IT systems. They also provide short-term backup power during grid interruptions before the backup generators take over for long-term outages. These systems contain batteries on their DC segment that have the potential to be utilized in the future for grid-supporting services, using their idle capacity that can absorb or inject power.
	
	\item Thermal storage and thermal inertia: Chilled water tanks and the facility’s own thermal mass could be utilized by the operators for precooling or changing the temperature set points, therefore allowing peak shaving and load shifting without any manipulation to the IT workload.
    
	\item Dynamic IT workload management: Studies show that around 60\% of IT or server tasks  \cite{cao2022data} are tolerant of delays. This flexibility can be exploited by shifting the workload temporally or spatially, therefore providing dynamic workload management  \cite{colangelo2026ai}.  
	
	\item Backup generators: During a period when the grid is stressed, the data center site can island itself and have the technical capability to even inject power back into the grid, especially during strenuous conditions such as restoration and synchronization.
	
\end{itemize}

Considering these attributes, data centers can be converted from firm loads into interruptible loads (fully or partially), enabling the electric utilities to tap into their existing grid headroom capacity for serving flexible data centers. Consequently, such flexibility help could accelerate the interconnection of AI data centers. For the flexible data centers to become operational, there are many nuances that the host electric utility and the data center enterprise need to thoroughly evaluate and mutually agree upon, including flexibility expectations, performance requirements, contractual obligations, and incentives.

We envision that large data centers will potentially play a larger role in grid real-time operation as they can be practically treated as microgrids. There are a host of services that these facilities may provide to the grid and act as a catalyst for decarbonization, for example, acting as nonwire alternative assets for peak shaving and load shifting (both temporal and spatial), congestion relief, short-term reserve for frequency and voltage response, helping with improved damping of oscillations, and intentional islanding during extreme events and helping with restoration for resilience. 

Most modern utility control rooms are equipped with the information and communication technologies and control infrastructure needed to implement these services. They are, namely, energy management systems for controlling transmission-connected assets and advanced distribution management systems and distributed energy resource management systems for managing the distribution-connected assets and coordination with distributed resources \cite{Luka}. However, the key enabler lies in the terms and conditions of business/commercial agreements between data centers and utilities on the response mechanism and specifications and the obligations from both sides, including the technical details such as the measurement information that will be used for determining market tariffs and payment processes. Furthermore, before data centers can safely become grid flexibility assets, there remain open questions to clarify the coordination among balancing authority, market operator, transmission owner and operator, and load entities with regards to responsibilities, data sharing, decision making, procedures, jurisdiction, and authorities to prevent any missteps or noncompliance.

\section{Closing Remarks}

Artificial intelligence is widely regarded as the cornerstone of the fourth industrial revolution, with large-scale data centers as the backbone driving this progress. To achieve sustainable growth of these facilities, data center enterprises, power utilities, and energy regulators need to come together and make adjustments to their processes, regulations, and standards to streamline their expedited interconnection and secure operation. We are optimistic that this article could serve as an informative basis for all parties involved and shed light on the opportunities and challenges that large AI data centers present for reevaluation and development of reliability standards, performance requirements, and market tariffs to prioritize the ratepayers and grid security and simultaneously paving the path for a sustainable AI boom.

\section*{Acknowledgment}

We thank Ramanathan Thiagarajan for his insightful comments. This work was in part funded by the Climate Innovation Collaboratory (CIC), an ongoing alliance between Deloitte and University of Colorado Boulder.

\end{document}